\documentclass[aps,prc,twocolumn,showpacs,showkeys,preprintnumbers,amsmath,amssymb,a4paper]{revtex4}
\usepackage{graphicx}
\usepackage{bm}
\usepackage{epsfig}

\newcommand{\be}{\begin{eqnarray}}
\newcommand{\ee}{\end{eqnarray}}

\def\bfg #1{{\mbox{\boldmath $#1$}}}

\begin{document}
\title{Elastic $\bar p d$ scattering and total $\bar p d$ cross sections reexamined}

\author{Yu.N.~Uzikov$^{1,2}$ and J.~Haidenbauer$^{3}$}

\affiliation{
$^1$Laboratory of Nuclear Problems, JINR, 141980 Dubna, Russia\\
$^2$Physics Department, Moscow State University, 119991 Moscow, Russia \\
$^3$Institute for Advanced Simulation and 
Institut f\"ur Kernphysik, Forschungszentrum J\"ulich,
D-52425 J\"ulich, Germany \\
}
\date{\today}

\begin{abstract}
We update our recent analysis of $\bar p d$ scattering, performed within 
the Glauber theory including the single- and double $\bar p N$ scattering 
mechanisms.  Specifically, now we consider also $\bar N N$ amplitudes from a new
partial-wave analysis of $\bar p p$ scattering data. 
Predictions for differential cross sections and the spin observables 
$A_y^d$, $A_y^{\bar p}$, $A_{xx}$, $A_{yy}$ are presented for antiproton beam
energies between 50 and 300 MeV.
Total polarized cross sections are calculated utilizing the optical theorem.
The efficiency of the polarization buildup for antiprotons in a storage ring is
investigated.
\end{abstract}


\maketitle
  
\section{Introduction}

Scattering of antiprotons off polarized nuclei can be used to produce
a beam of polarized antiprotons by exploiting the so-called spin-filtering
effect \cite{barone}. The PAX Collaboration intends to utilize 
scattering of antiprotons off a polarized $^1$H target in
rings \cite{frank05} as the basic source for antiproton polarization 
buildup at an upgrade of the FAIR facility in Darmstadt. 
In view of the limited information on the spin dependence of the $\bar pN$ 
force, the interaction of antiprotons with a polarized deuteron is also
of interest for the issue of the antiproton polarization buildup.

In a recent paper \cite{UH2012} we considered $\bar p d$ scattering within 
the Glauber theory of multi-step scattering \cite{GlauberFranco}, 
taking into account the full spin-dependence of the elementary $\bar p N$ 
scattering amplitudes.
Predictions for various observables were given for antiproton beam energies 
from 50 to 300 MeV employing $\bar p N$ amplitudes generated from 
$\bar N N$ potentials developed by the J\"ulich Group \cite{Hippchen,Mull1,Mull2}. 
Specifically, the $\bar pN$ amplitudes were taken from the models 
A(BOX) introduced in Ref.~\cite{Hippchen} and D described in 
Ref.~\cite{Mull2}. Both models provide a very good overall reproduction of 
the low- and intermediate energy $\bar N N$ data as documented in those works.
On the other hand, there are clearly visible deficiencies in the 
description of spin-dependent observables like the analyzing powers, for
elastic $\bar p p$ scattering but in particular for the 
$\bar p p\to \bar n n$ reaction\cite{Mull2}. 

\begin{figure}
\mbox{\epsfig{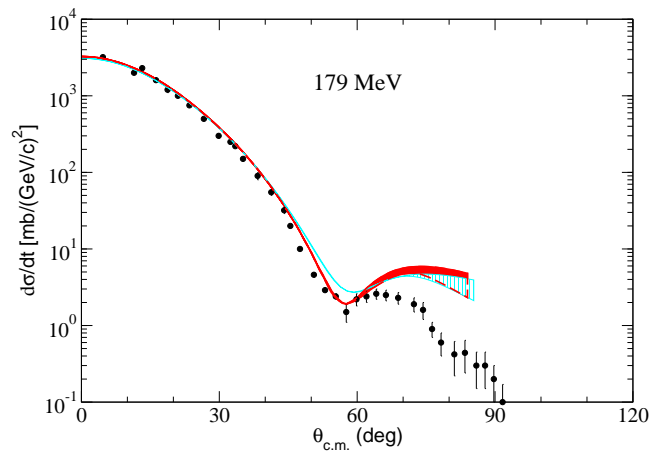}}
\caption{(Color online) Differential cross section of elastic ${\bar p} d$
 scattering at 179~MeV versus the c.m. scattering angle.
 Results are shown for the $\bar NN$ amplitudes of Ref.~\cite{Timmermans2}
 (cyan/hatched) and of the J\"ulich model D (red/black). 
 The bands represent the sensitivity to variations of the large-angle tail 
 of the $\bar p N$ amplitudes as discussed in the text. 
 The data points are taken from Ref. \cite{bruge}.}
\label{pd179q2}
\end{figure}
 
\begin{figure*}[t]
\mbox{\epsfig{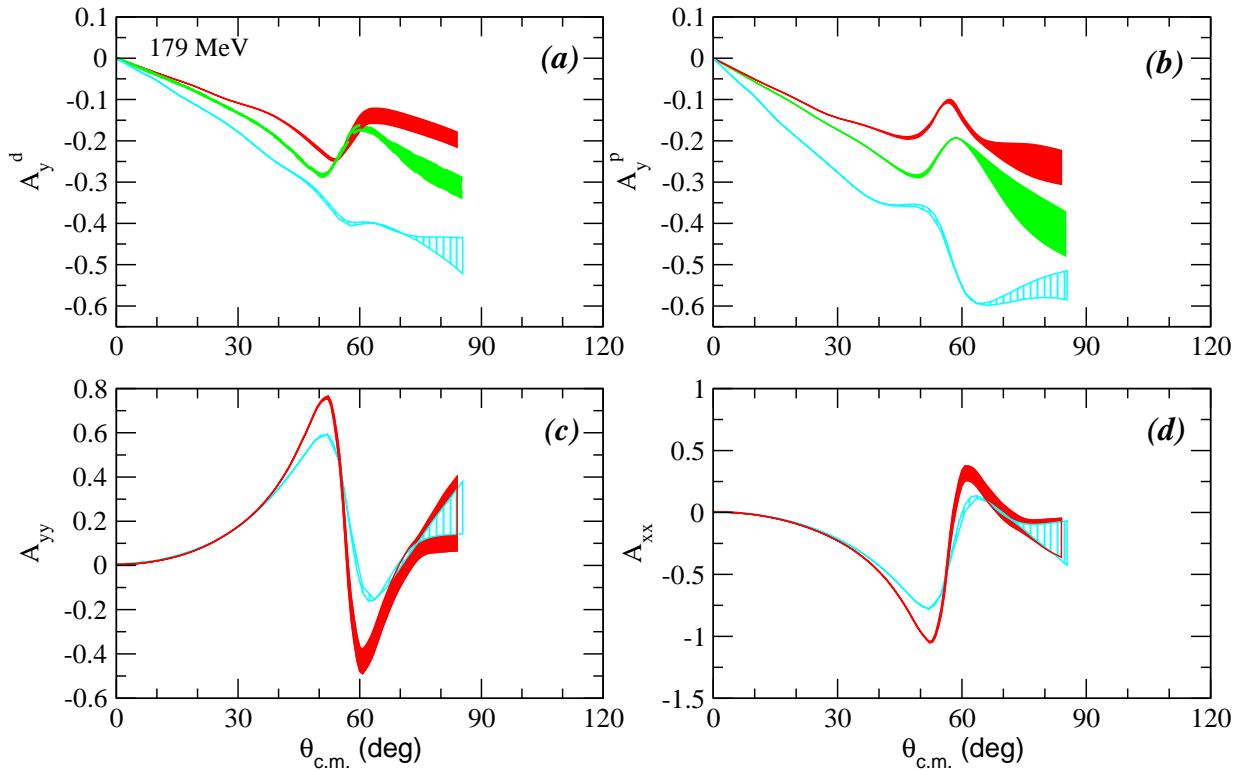}}
\caption{(Color online) Spin observables of elastic ${\bar p}d$ scattering at 
179 MeV versus the c.m. scattering angle:
$A_y^d$ (a), $A_y^{\bar p}$ (b), $A_{yy}$ (c), and $A_{xx}$ (d).
Results are shown for the $\bar NN$ amplitudes of Ref.~\cite{Timmermans2}
(cyan/hatched), and of the $\bar NN$ models A (green/grey) and
 D (red/black).
 The bands represent the sensitivity to variations of the large-angle tail 
 of the $\bar p N$ amplitudes as discussed in the text.
 }
\label{ay179}
\end{figure*}

While we were preparing our work \cite{UH2012} an updated version of the Nijmegen 
partial-wave analysis was presented by Zhou and 
Timmermans \cite{Timmermans2}. For this new analysis concrete  
values of the resulting phase shifts and inelasticities are provided in 
the publication so that we can reconstruct the corresponding $\bar NN$ 
amplitudes and we can employ them for a calculation of $\bar p d$ 
scattering within the Glauber theory. As demonstrated in \cite{Timmermans2}
the $\bar NN$ amplitudes based on those phase-shift parameters
yield an excellent description of the experimental data included in the database 
of the analysis. Thus, they constitute definitely the best and most reliable 
representation of the $\bar NN$ interaction, and specifically of its spin
dependence, that we have at hand at the moment. 
Therefore, it is instructive to investigate the implications of those
amplitudes on the various $\bar p d$ scattering observables that we
considered in our recent paper \cite{UH2012}. Of particular interest are, of
course, spin observables such as $A_y^d$, $A_y^{\bar p}$, $A_{xx}$, $A_{yy}$.
Equally interesting are predictions for the efficiency of the polarization buildup 
for antiprotons in a storage ring for scattering off a deuteron target. 
Those are the parameters relevant for the planned experiments of the PAX 
Collaboration and it is rather helpful for the preparations of a future experiment
to have values at one's disposal that are as well-founded as possible. 
Corresponding results are presented in this Brief Report.   
In the following we use the abbreviation ``ZT'' when we refer to the
amplitudes of Zhou and Timmermans \cite{Timmermans2}. 

Note that $\bar p d$ scattering was also considered in Ref.~\cite{salnikov} 
utilizing, however, the results from the old Nijmegen $\bar pp$ partial-wave 
analysis \cite{Timmermans} from 1994. 

\section{Results and discussion}

We study $\bar p d$ scattering within the Glauber theory based on the 
single- and double $\bar p N$ scattering mechanisms.
The full spin dependence of the elementary $\bar p N$ scattering amplitudes is
taken into account and both the $S$- and $D$-wave components of the deuteron
are considered. Details of the formalism can be found in Refs.~\cite{UH2012,ujh2009},
where we also provide definitions of the considered $\bar p d$ 
observables in terms of the twelve invariant amplitudes that arise
for spin 1/2+1 scattering. 

The $\bar N N$ amplitudes of the new analysis of Zhou and Timmermans
are reconstructed from the phase shifts and inelasticity parameters
as given in their tables VIII-X \cite{Timmermans2}. Those values are
obtained under the assumption of isospin symmetry and, therefore, 
match our study where likewise isospin symmetry is imposed for the
hadronic amplitude. 
Since only partial waves up to a total angular momentum of $J=4$ are
listed we considered two options for supplementing the contributions
from higher partial waves, namely (a) one-pion exchange and (b) higher
partial waves predicted by the J\"ulich model A. It turned out that
there is very little difference in the resulting $\bar N N$ amplitudes,
at least up to $p_{lab} = 800$ MeV/c, which is the highest momentum
considered in our study. Actually, even in a test calculation of
differential cross sections and analyzing powers based on the phase 
shifts at $900$ MeV/c with our reconstructed amplitudes we obtained 
nice agreement with the results for $\bar pp$ elastic and $\bar pp\to \bar nn$ 
charge-exchange scattering (at $860\approx886$ MeV/c) displayed in
Ref. \cite{Timmermans2}.

Following our previous work we use a Gaussian ansatz for representing 
the amplitudes generated from the $\bar NN$ phase-shift parameters
of Ref. \cite{Timmermans2} in analytical form. Again, we aim at an 
excellent reproduction of the original amplitudes over the whole angular
range. In the $\bar pd$ calculation we introduce a cutoff that suppresses 
the amplitudes in the backward hemisphere as described in \cite{UH2012} 
and we vary the cutoff angle in the $\bar p d$ calculations. 
The bands in the figures represent the variation of the $\bar p d$ 
observables due to the procedure described above.
As argued in \cite{UH2012}, we regard these bands as a sensible guideline 
for estimating the angular region where the Glauber theory is able to 
provide solid results for a specific observable and where this approach 
starts to fail. We want to remind the reader that 
contributions from large angles are in contradiction with the basic
approximations underlying the Glauber model and, thus, any sizable 
influence from them undoubtedly marks the breakdown of this approach.

\begin{figure}[hbt]
\mbox{\epsfig{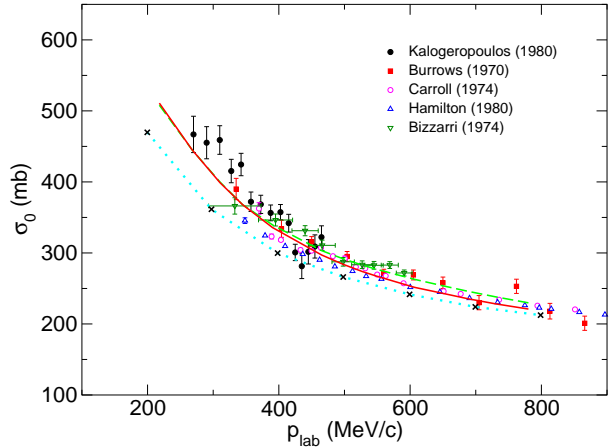}}
\caption{(Color online) Total unpolarized $\bar p d$ cross section versus the antiproton 
laboratory momentum.
Results are shown for the $\bar NN$ amplitudes of Ref.~\cite{Timmermans2} 
(dotted line) and for the $\bar NN$ models A (dashed line) and D (solid line).
Data are taken from Refs.  \cite{HERA,Carroll,Hamilton}.
 }
\label{totsigm0}
\end{figure}

First let us consider the differential cross section at 179 MeV 
where data are available \cite{bruge}, see Fig.~\ref{pd179q2}. 
Our Glauber calculation describes the first diffractive peak quite well 
- for amplitudes taken from the $\bar pp$ partial-wave analysis as
well as for those generated from the $\bar NN$ model D of
the J\"ulich Group. 
(The result for the former is based on the amplitudes at 175.3 MeV 
($p_{lab}= 600$ MeV/c) listed in Ref.~\cite{Timmermans2}.) 
Model D also explains the first minimum in the differential cross section,
located at $q^2\approx 0.12-0.13$ (GeV/c)$^2$
(i.e. $\theta_{c.m.}\approx 55^\circ$), and the onset of the second
maximum whereas here the ZT $\bar NN$ amplitudes lead to an overestimation. 
The obvious strong disagreement with the data at larger transferred momenta, 
$q^2> 0.15$ (GeV/c)$^2$,
corresponding to $\theta_{c.m.}>60^\circ$, lies already in the region where
the increase in the band widths indicates that our Glauber results are not
reliable anymore, cf. the corresponding discussions in Ref.~\cite{UH2012}. 

In case of the vector analyzing powers $A_y^{\bar p}$ and $A_y^d$ our investigation
\cite{UH2012} had indicated a strong model dependence. Thus, it is not surprising
that the corresponding predictions for the ZT amplitudes differ from those of
the J\"ulich models as can be seen in Fig. \ref{ay179}. Indeed, the ZT amplitudes 
yield significantly larger values for those observables. 
The tensor analyzing powers $A_{xx}$ and $A_{yy}$ were found to be much less 
sensitive to differences in the $\bar N N$ amplitudes because these observables 
are dominated by the spin-independent amplitudes. 
For the ZT amplitudes we obtain results that exhibit an angular dependence 
that is very similar to the one predicted by the J\"ulich models. 
Indeed the results almost coincide with those for model A and, therefore,
we do not display the latter for reasons of clarity.

\begin{figure}[hbt]
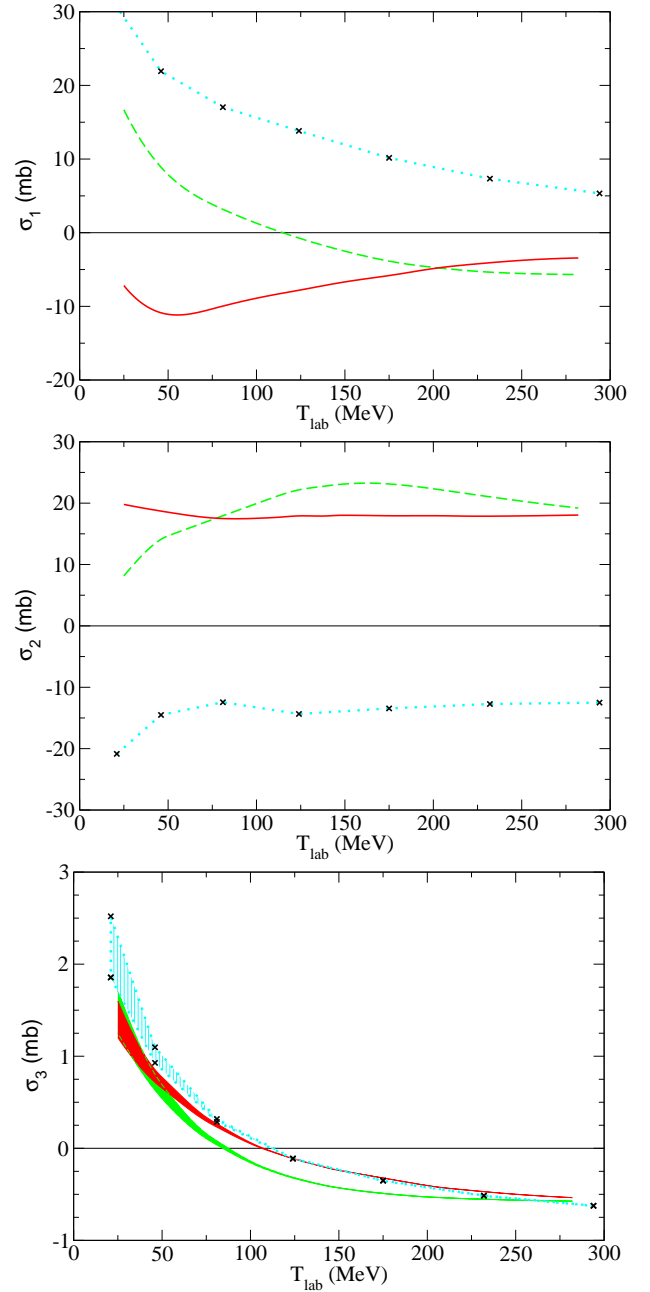

\mbox{\epsfig{figure=sigma1n.eps,height=0.24\textheight, clip=}}
\mbox{\epsfig{figure=sigma2n.eps,height=0.24\textheight, clip=}}
\mbox{\epsfig{figure=sigma3n.eps,height=0.24\textheight, clip=}}
\caption{(Color online) Total $\bar p d$ cross section $\sigma_1$,
$\sigma_2$, and $\sigma_3$ versus 
the antiproton kinetic energy in the laboratory system. 
Same description of curves as in Fig. \ref{totsigm0}. 
 }
\label{totsigm1}
\end{figure}

The total $\bar p d$ cross section is defined by
 \cite{ujh2009}
\begin{equation}
\label{totalspin}
\sigma=\sigma_0+\sigma_1{\bf P}^{\bar p}\cdot {\bf P}^d+
 \sigma_2 ({\bf P}^{\bar p}\cdot {\bf \hat  k}) ({\bf P}^d\cdot {\bf \hat k})+
\sigma_3 P_{zz},
\end{equation}
where ${\bf \hat k}$ is the unit vector in the direction of the antiproton beam,
 ${\bf P}^{\bar p}$ ($ {\bf P}^d$) is the polarization vector of the antiproton (deuteron),
 and $P_{zz}$ is the tensor polarization of the deuteron ($OZ||{\bf \hat k}$).
Corresponding results are presented in Figs. \ref{totsigm0} and
\ref{totsigm1}. It is obvious that the unpolarized cross section $\sigma_0$ 
(Fig.~\ref{totsigm0}) based on the ZT amplitudes is somewhat smaller 
than those for the J\"ulich $\bar NN$ models. It is also below the 
bulk of the experimental data \cite{HERA,Carroll,Hamilton}.
A closer inspection revealed that this difference is mainly due to 
a qualitative difference in the magnitude of the isospin $I=1$ amplitude. 
In the J\"ulich models $\sigma_{\bar pn} \propto |T_{I=1}|^2$ is about
10-15 \% larger than the result we obtain for the $I=1$ amplitude of 
Ref.~\cite{Timmermans2}. 
The $\bar pp$ and $\bar pp \to \bar nn$ cross sections are given
by $\sigma_{\bar pp} \propto |(T_{I=0}+T_{I=1})/2|^2$ and
$\sigma_{\bar pp \to \bar nn} \propto |(T_{I=0}-T_{I=1})/2|^2$, 
respectively, so that interferences between the $I=0$ and $I=1$
amplitudes play a role and the (absolute) size of the isospin 
amplitudes is not so tightly constrained. 
However, the $\bar p d$ scattering cross section is given 
(in the simple impulse approximation \cite{ujh2009}) by 
$\sigma_{\bar pd} \propto (|(T_{I=0}+T_{I=1})/2|^2+|T_{I=1}|^2)$. 
Note that the calculation for the ZT amplitudes was done at those energies
(marked by ``${\bf x}$'' in the figures) for which the values are listed 
in \cite{Timmermans}.
To guide the eye we connected those points with a dotted line.  

\begin{figure}[hbt]
\mbox{\epsfig{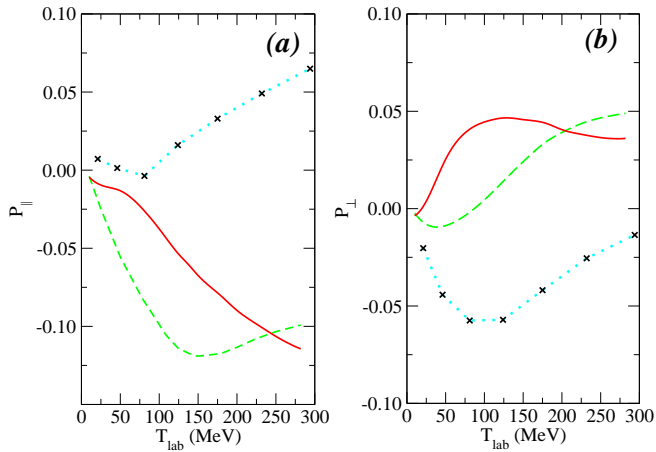}}
\caption{(Color online) Dependence of the (a) longitudinal ($P_{||}$) 
and (b) transversal ($P_{\perp}$) polarization on the beam energy. 
Same description of curves as in Fig. \ref{totsigm0}. 
The acceptance angle is 20 mrad.
 }
\label{fmerit}
\end{figure}

Predictions for the spin-dependent $\bar p d$ cross section $\sigma_1$,
$\sigma_2$, and $\sigma_3$ are shown in Fig.~\ref{totsigm1}. 
Again, as for the analyzing powers discussed above, we see sizable 
differences in the results based on the ZT amplitudes to the ones obtained 
from the $\bar NN$ amplitudes of the J\"ulich models. Specifically,
those cross sections are larger ($\sigma_1$) or even of oppositive
sign ($\sigma_2$). Only the tensor-polarized cross section
$\sigma_3$ turns out to be similar for all three considered $\bar NN$ 
amplitudes.

The quantity relevant for the efficiency of the polarization buildup is
the polarization degree $P_{\bar p}$ at the beam life time $t_0$ \cite{mss}. 
With our definition of $\sigma_1$ and $\sigma_2$ \cite{UH2012} this quantity is 
given by
\begin{eqnarray}
\nonumber
&&P_{\bar p} (t_0)=-2P_T\frac{\sigma_1}{\sigma_0}, \ {\rm if} \ {\bfg \zeta}\cdot {\bf \hat k}=0, \\
&&P_{\bar p} (t_0)=-2P_T\frac{\sigma_1+\sigma_2}{\sigma_0}, \ {\rm if} \
 |{\bfg \zeta}\cdot {\bf \hat k}|=1 \ ,
\label{RL}
\end{eqnarray}
where the unit vector ${\bfg \zeta}$ is directed along the target polarization vector $P_T$.
Results for the transversal polarization $P_{\perp}$ (${\bfg \zeta}\cdot {\bf \hat k}=0$)
and for the longitudinal polarization $P_{||}$ (${\bfg \zeta}\cdot {\bf \hat k}=1$)
are shown in Fig.~\ref{fmerit} for $P_T = P^d = 1$. 
Obviously, there are sizable differences in the predicted values for the considered
$\bar NN$ amplitudes. But the overall magnitude -- which is decisive for the 
polarization buildup -- is comparable and in the order of 5-10 \% in the energy region 
considered. With regard to the calculation based on the $\bar NN$ amplitudes from the
partial-wave analysis \cite{Timmermans2} we want to point out that our 
results for the polarization degree for a deuterium target are smaller
than those for a hydrogen target, found to be around 20 \% in Ref.~\cite{Timmermans3} 
for the acceptance angle of 20 mrad considered by us.
Moreover, they are much smaller than the large values of around 30 \% reported 
in Ref.~\cite{salnikov} in a $\bar p d$ calculation that utilizes the old Nijmegen 
$\bar NN$ partial-wave analysis \cite{Timmermans}.

\noindent
{\bf Acknowledgements:}
 This work was supported in part by the Heisenberg-Landau program.


\end{document}